\documentclass[useAMS,usenatbib]{mn2e}

\usepackage{color}
\usepackage{amsmath}
\usepackage{graphicx}

\def\be{ \begin{equation} }

\def\ee{\end{equation}}

\def\ba#1\ea{\begin{align}#1\end{align}}

\newcommand{\vs}{\nonumber\\}

\newcommand{\refeq}[1]{Eq.~(\ref{eq:#1})}          
          
\newcommand{\reffig}[1]{figure~\ref{fig:#1}}          
\newcommand{\refFig}[1]{Figure~\ref{fig:#1}}          
\newcommand{\refsec}[1]{section~\ref{sec:#1}}

\newcommand{\Om}{\Omega_m}

\newcommand{\OL}{\Omega_\Lambda}

\renewcommand{\d}{\delta}
\newcommand{\tOm}{\tilde{\Omega}_m}
\newcommand{\tOK}{\tilde{\Omega}_K}

\newcommand{\tOL}{\tilde{\Omega}_\Lambda}

\def\tt{\tilde t_{0}}

\title[Separate Universe Simulations]{Separate Universe Simulations}
\author[C. Wagner et al.]{C. Wagner$^{1}$\thanks{E-mail:
cwagner@mpa-garching.mpg.de}, F. Schmidt$^{1}$, C.-T. Chiang$^{1}$ and E. Komatsu$^{1,2}$\\
$^{1}$Max-Planck-Institut f\"ur Astrophysik, Karl-Schwarzschild-Str. 1, 85741 Garching, Germany\\
$^{2}$Kavli Institute for the Physics and Mathematics of the Universe, Todai Institutes for Advanced Study, \\
University of Tokyo, Kashiwa, Japan 277-8583 (Kavli IPMU, WPI)}

\begin{document}


\pagerange{\pageref{firstpage}--\pageref{lastpage}} \pubyear{2014}

\maketitle

\label{firstpage}

\begin{abstract}
The large-scale statistics of observables such as the galaxy density are
chiefly determined by their dependence on the local coarse-grained matter
density.  This dependence can be measured directly and efficiently in N-body 
simulations by using the fact that a uniform density perturbation with respect 
to some fiducial background cosmology is equivalent to modifying the background 
and including curvature, i.e., by simulating a ``separate universe''.  
We derive this mapping to fully non-linear order, and provide a step-by-step
description of how to perform and analyse the separate universe simulations.  
This technique can be applied to a wide range of observables.  
As an example, we calculate the response of the non-linear matter
power spectrum to long-wavelength density perturbations, which corresponds
to the angle-averaged squeezed limit of the matter bispectrum and higher
$n$-point functions.  Using only a modest simulation volume, we obtain
results with percent-level precision over a wide range of scales.
\end{abstract}


\section{Introduction}
\label{sec:intro}

The question of how some physical quantity responds to a uniform change
in the matter density, expressed as overdensity $\delta_\rho$, arises in
several contexts in cosmology. The most prominent example is probably
the bias of galaxies or other tracers of the matter distribution. In the
local bias prescription of \cite{fry/gaztanaga:1993} the overdensity of
the tracers is expanded in terms of the matter overdensity as $\delta_g
= b_1 \delta_\rho + \frac12 b_2 \delta_\rho^2 + ...$, where $b_1$ is the linear bias and $b_2$ and higher orders are called non-linear bias parameters. Another example, which we consider in detail below, is the analogous expansion of the non-linear matter power spectrum $P(k)$.  The $k$-dependent coefficients, which we refer to as \emph{power spectrum response} functions, describe how the non-linear growth of matter perturbations is modified by a change of the background density.  
The response to a change in the matter density is also of interest for many other quantities, e.g., halo profiles, the number density of voids, weak lensing shear and convergence, and so on.

Usually, an indirect approach is used to study this kind of question. The halo bias, for example, is usually measured indirectly from clustering statistics: the linear bias $b_1$ from the power spectrum on large scales, $b_2$ from the bispectrum on large scales \citep[e.g.,][]{nishimichi/etal:2007,baldauf/etal:2012} or from cumulants of the smoothed halo and matter fields \citep[e.g.,][]{angulo/baugh/lacey:2008,manera/gaztanaga:2011}.

In this Letter, we show how one can efficiently apply a change of the matter density to N-body simulations, by using the fact that a uniform density perturbation on an FRW background is equivalent to a different (curved) FRW background. 
The idea of treating a patch of the Universe as a separate universe goes back to \citet{lemaitre:1933} and has been used in many calculations since then.
However, the implementation of the separate universe picture in N-body simulations is relatively new and was only worked out to the lowest order in $\delta_\rho$  \citep{mcdonald:2003, sirko:2005}.\footnote{See \citet{gnedin/kravtsov/rudd:2011} for an alternative approach to incorporating a uniform density perturbation.} Recently, this technique was used to measure the linear response of the non-linear power spectrum from simulations \citep{li/hu/takada:2014}. 
Here, we will not assume that $\delta_\rho$ is small.  

This approach offers several advantages over the widely used indirect methods:  first, it isolates the effect of a uniform density perturbation from those of tidal fields and density gradients.  Second, by allowing for an (in principle) arbitrarily large uniform density perturbation, one can derive not only the first order but all higher order responses of the desired observable.  Finally, we can largely remove the noise due to cosmic variance by performing simulations with fixed initial phases, allowing for a very small statistical error from a modest simulation volume.  Further, one such set of simulations can be used for many of the applications mentioned above.
These key advantages, together with the simple step-by-step prescription 
presented here, should make this approach highly useful for a wide range
of applications in large-scale structure.  As an example, we demonstrate the technique by applying it to the response functions of the matter power spectrum,
showing that we can achieve percent-level measurements of the squeezed limit
of the non-linear matter bispectrum, trispectrum, and 5-point function, on
scales ranging from $k \simeq 0.01$ to $1\ h\,{\rm Mpc}^{-1}$.   

\vspace{-0.4cm}
\section{Separate Universe}
\label{sec:sep_univ}
The idea of the separate universe technique is to absorb the overdensity
$\delta_\rho$ into the background density of a modified cosmology
$\tilde\rho(t)$ as \citep{sirko:2005,baldauf/etal:2011,sherwin/zaldarriaga:2012,li/hu/takada:2014}
\ba
\rho(t) \left[ 1 +\delta_\rho(t)\right] = \tilde\rho(t) \,. \label{eq:absorb}
\ea
Thus, instead of embedding the region with overdensity $\delta_\rho$ in a fiducial  background universe, one considers it as a separate universe with an altered cosmology. In this section we derive the mapping between the cosmological parameters of the fiducial and modified cosmology as a function of the linearly extrapolated present-day overdensity $\delta_{L0}=\delta_\rho(t_i) D(t_0)/D(t_i)$, where $D$ is the linear growth function of the fiducial cosmology, $t_0$ the present time, and $t_i$ an early time at which $\delta_\rho$ is still small. Throughout, we will not assume $\delta_{L0}$ to be small.

Expressed in terms of the standard cosmological parameters, i.e.~$\rho(a\rm{=}1)=\Omega_m \frac{3H_0^2}{8\pi G}$ and $H_0=h\,100\,\rm{km\, s}^{-1}\rm{Mpc}^{-1}$, \refeq{absorb} becomes
\ba
\frac{\Omega_m h^2}{a^3(t)} \left[ 1 +\delta_\rho(t)\right] = \frac{\tilde\Omega_m \tilde h^2}{\tilde a^3 (t)} \,,
\ea
where we used $\ \tilde{}\ $ to denote quantities in the modified cosmology. 
For the fiducial cosmology, we adopt the standard convention for the scale factor $a(t_0)=1$. In contrast, for the modified cosmology, it is convenient to choose $\tilde a(t\to 0) = a$ as $\delta_\rho(t\to 0)=0$. These conventions lead to
\be 
\Omega_m h^2=\tilde \Omega_m \tilde h^2\,.
\label{eq:omh2}
\ee
Introducing $\delta_a(t)$ by $\tilde a(t)=[1+\delta_a(t)] a(t)$, we find
\be
1 + \d_\rho(t) = [1 + \d_a(t)]^{-3}\,.
\label{eq:dadrho}
\ee
In the following we use the first and second Friedmann equations of the two cosmologies to derive a differential equation for $\delta_a$ (and thereby also for $\delta_\rho$). 
The Friedmann equation for $a(t)$ is
\be
H^2(t) = \left(\frac{\dot a}a\right)^2 
= \frac{8\pi G}3 \left[\rho(t) + \rho_X(t) \right]\,,
\label{eq:Fr1}
\ee
where $\rho_X(t)$ denotes the Dark Energy (DE) density and we have assumed a flat
fiducial cosmology for simplicity.  The same equation, but including 
curvature $\tilde K$ and modifying the densities, holds for $\tilde a(t)$:
\be
\tilde H^2(t) = \left(\frac{\dot{\tilde a}}{\tilde a}\right)^2 = 
\frac{8\pi G}3 \left[\tilde\rho(t) + \tilde\rho_X(t) \right]
 - \frac{\tilde K}{\tilde a^2(t)}\,,
\label{eq:Fr1t}
\ee
Further, we have the second Friedmann equation,
\be
\frac{\ddot a}{a} = - \frac{4\pi G}3 \left[ \rho(t) + \rho_X(t) + 3 p_X(t)\right],
\label{eq:Fr2}
\ee
and correspondingly
\be
\frac{\ddot{\tilde a}}{\tilde a} = - \frac{4\pi G}3 \left[ \tilde\rho(t) 
+ \tilde\rho_X(t) + 3 \tilde p_X(t) \right]\,.
\label{eq:Fr2t}
\ee
If DE is not a cosmological constant, there are also perturbations in the DE fluid $\delta\rho_X\equiv \tilde\rho_X - \rho_X$ and $\delta p_X\equiv \tilde p_X - p_X$.
In order for the separate universe approach to work, matter and DE have to be comoving and follow geodesics of the FRW metric. 
Since this requires negligible pressure gradients, the approach is only applicable to density perturbations with wavelengths $2\pi/k$ that are much larger than the sound horizon of DE: $k\ll H_0/|c_s|$, where the sound speed is defined by $c_s^2=\delta p_X / \delta \rho_X$ (see also \citet{creminelli/etal:2010}). Hence, the size of the simulation box has to be much larger than the DE sound horizon.\footnote{In order to simulate DE consistently inside the simulation box, the simulation code needs to take DE perturbations and resulting pressure forces into account. However, this is also the case for the fiducial cosmology and is not a consequence of the separate universe picture.}
For simplicity, we assume from now on that DE is just a cosmological constant $\Lambda$. In this case $\tilde\rho_\Lambda=\rho_\Lambda$ and $\tilde p_\Lambda= p_\Lambda = -\rho_\Lambda$.

Using the definition of $\d_a(t)$
we can write 
\ba
\label{eq:dH}
\tilde H  =\:& \frac{\left[1 + \d_a \right] \dot a + \dot\d_a\,a}{[1+\d_a] a} = H + \frac{1}{1+\d_a} \dot\d_a
\\
\frac{\ddot{\tilde a}}{\tilde a} =\:& 
\frac{\left[1 + \d_a \right] \ddot a 
+ 2\dot\d_a \dot a 
+ \ddot\d_a\,a}{(1+\d_a) a}
= \frac{\ddot a}{a} +  \frac{1}{1+\d_a}\left[\ddot\d_a + 2 H \dot\d_a\right]
\,.\nonumber
\ea
Inserting the second Friedmann equations [\refeq{Fr2} and \refeq{Fr2t}] into the above equation  yields an ordinary differential equation
for the perturbation to the scale factor:
\ba
&\ddot\d_a + 2 H \dot\d_a = - \frac{4\pi G}3  \rho(t) 
\left([1 + \d_a(t)]^{-3} - 1 \right) \left[1 + \d_a(t) \right]\,, 
\label{eq:daNL}
\ea
where we have used \refeq{dadrho} to express $\tilde \rho(t)$ in terms of $\delta_a$ and $\rho(t)$.  
One can also transform this equation to $\d_\rho(t)$, yielding 
\be
\ddot\d_\rho + 2 H \dot\d_\rho - \frac43 \frac{\dot\d_\rho^2}{1+\d_\rho}
= 4\pi G \rho\, (1+\d_\rho) \d_\rho\,.
\label{eq:drhoNL}
\ee
When linearising this equation in $\d_\rho$, one obtains the equation for
the linear growth factor.  More generally, \refeq{drhoNL}
is exactly the equation for the interior density of a spherical
tophat perturbation in a $\Lambda$CDM universe (see e.g. App.~A of \cite{HPMhalopaper}).

The difference of the modified and fiducial first Friedmann 
equations yields the curvature $\tilde K$:
\ba
\frac{\tilde K}{a^2(t)}  =\:& \frac{8\pi G}{3} \rho(t) 
\left([1 + \d_a]^{-1} - [1+\d_a]^2 \right)\vs
\:& - 2 H (1+\d_a) \dot\d_a - \dot\d_a^2\,,
\label{eq:Ktilde}
\ea
where we used \refeq{dH} to express $\tilde H$.  In order to be a valid Friedmann model, the curvature $\tilde K$ has to be conserved. Using \refeq{daNL} and the continuity equation, it can readily be shown that this is indeed the case.  
Thus, we can evaluate \refeq{Ktilde} at an early time $t_i$, when the perturbation $\d_a$ is infinitesimal and the Universe is in matter domination.  
We then have $H^2=H_0^2 \Om a^{-3}$, $\dot\d_a=H\d_a$, and $\d_a=-\d_\rho/3$, with which we derive
\be
\frac{\tilde K}{H_0^2} = \frac53 \frac{\Om}{a(t_i)} \d_\rho(t_i)\,.
\ee
Alternatively, using the linear growth
factor normalized such that $D(t_i)=a(t_i)$, we can write
\be
\frac{\tilde K}{H_0^2} = \frac53 \frac {\Om}{D(t_0)}  \d_{L0}\,.
\label{eq:deltaKL}
\ee

Now let us derive the parameters of the modified cosmology. They are defined through the Friedmann equation at time $\tt$ where  $\tilde a(\tilde t_0)=1$. This is given by \refeq{Fr1t},
\be
\tilde H^2(t) = \tilde H_0^2 \left( \tOm\,\tilde a^{-3}(t)
+ \tOL  + \tOK \tilde a^{-2}(t) \right) \\
\ee 
with
\ba
\tilde H_0 \equiv \:&  \tilde H(\tt) &
\tOK \equiv\:&  - \frac{\tilde K}{\tilde H_0^2}  \vs
\tOm \equiv \:& \frac{8\pi G}{3\tilde H_0^2} \tilde\rho(\tt) &
\tOL \equiv \:& \frac{8\pi G}{3\tilde H_0^2} \rho_\Lambda \,.
\ea
Defining $\d_H$ through $\tilde H_0 = H_0 [ 1 + \d_H]$ 
and using \refeq{omh2}, we obtain
\be
\tOm = \Om [ 1 + \d_H]^{-2}; \quad
\tOL = \OL [ 1 + \d_H]^{-2}\,.
\ee
Finally, in order to derive $\delta_H$ we can make use of the Friedmann equation at $\tt$, which yields
\be
\tOK = -\frac{\tilde K}{\tilde H_0^2} = 1 - \tOm - \tOL = 1 - (1+\d_H)^{-2}\,,
\ee
where we have used $\Om + \OL = 1$ (since the fiducial cosmology is flat).  
Since $\tilde K$ is given by \refeq{deltaKL}, we
can use this relation to solve for $\d_H$:
\be
\d_H = \left(1-\frac{\tilde K}{H_0^2} \right)^{1/2} - 1\,.
\ee
There is no solution if $\tilde K/H_0^2 \geq 1$, or equivalently
$\d_{L0} \geq ( \frac53 \frac {\Om}{D(t_0)}  )^{-1}$.  This is because for such a large
positive curvature, the universe reaches turnaround at or before $\tilde a = 1$.
This is not a physical problem, it is merely not possible
to parametrize such a cosmology in the standard convention.  
For practical applications, smaller values of $\d_{L0}$ are in any case sufficient.  

When we run separate universe simulations,
we want to output the data at some physical time $t_{\rm out}$, which in
N-body codes is usually specified by the scale factor as $a(t_{\rm
out})=a_{\rm out}$. Therefore we need to determine the corresponding
scale factor in the modified cosmology as $\tilde a(t_{\rm out})=a_{\rm out}\left[1+\delta_a(t_{\rm out})\right]$. One way to compute $\delta_a(t_{\rm out})$ is to solve \refeq{daNL} numerically. Alternatively, we can simply determine $\delta_a(t_{\rm out})$ by requiring that the time given by the integral $t_{\rm out}=\int_0^{a_{\rm out}} da/\left[aH(a)\right]$  is the same in both cosmologies, i.e.,
\be 
\int_0^{a_{\rm out}} \frac{da}{aH(a)} = \int_0^{a_{\rm out}\left[1+\delta_a(t_{\rm out})\right]} \frac{d\tilde a}{\tilde a \tilde H (\tilde a)}\,.
\ee

\vspace{-0.4cm}
\section{N-body simulations}
\label{sec:nbody}

In order to generate the initial conditions for N-body simulations of the 
modified cosmologies, we need the linear power spectrum at the initial 
redshift.  The initial power spectrum has to be generated for the cosmology
$\{\tOm,\tOL,\tilde H_0\}$ with the same amplitude of the primordial scalar 
curvature perturbations $\mathcal{A}_s$ as for the fiducial cosmology.  
Since the transfer function only involves the physical matter and radiation 
densities quantified by $\tOm \tilde H_0^2$ and so on, it is the same in 
the modified and fiducial cosmologies.  Therefore the linear power spectra 
differ only through the difference in the linear growth.   
We take the power spectrum of the fiducial cosmology at $z=0$, in our case obtained from CAMB \citep{camb}, and rescale it by 
$\left[ \tilde D(\tilde a_i) \tilde D(\tilde a=1)/D(a=1)\right]^2$, where 
$\tilde D$ is the linear growth function of the modified cosmology and 
$\tilde a_i$ the scale factor for which the initial conditions are generated.\footnote{
Note that, in order to recover the correct linear power spectrum  at
low redshifts, we compute the growth functions ($D$ and $\tilde D$) without taking 
into account radiation.  This is because N-body codes do not include the effect of cosmological radiation. 
In our procedure, we also neglect the effect
of curvature on the transfer function at very low wavenumbers $k \sim \sqrt{|K|}$, since terms of similar order are neglected in the Poisson equation used in N-body codes.  For sub-horizon box sizes these effects are entirely negligible.}  Next we generate a Gaussian realization of the density field following the initial power spectrum.
The positions and velocities of the particles are then determined by using second-order Lagrangian perturbation theory (2LPT).

Usually in N-body codes, one specifies the box size in comoving units of $h^{-1}~{\rm Mpc}$ of the corresponding cosmology. 
Let us denote the comoving box size of the fiducial cosmology and of the modified cosmology in the above units by $L$ and $\tilde L$, respectively. 
The physical size of the boxes in units of Mpc is then given by $a(t) L/h$ and $\tilde a(t) \tilde L/\tilde h$.
In order to reduce the sample variance when comparing simulations with different overdensities $\delta_{L0}$, it is desirable that the simulations start from the same realization of the initial density field. However, there are two options in doing this \citep{li/hu/takada:2014}: Either the random fluctuations coincide on comoving scales at all times or the random fluctuations coincide on physical scales at only one specific time. In the former case one sets $\tilde L/ \tilde h=L/h$, while in the latter one requires $\tilde a(t_{\rm out})\tilde L/ \tilde h=a(t_{\rm out})L/h$, which implies that for each output time $t_{\rm out}$ a different value for $\tilde L$ is needed (when keeping $L$ fixed) and hence a new simulation has to be performed.
Which of the two options is better depends on the question one tries to answer. If one wants to single out the effect of the overdensity $\delta_{L0}$ on the growth of density modes with comoving scale $k$, then matching the random fluctuations on comoving scales is appropriate. If one instead is interested in the full effect of $\delta_{L0}$ including the dilation of scales due to the difference in the scale factors ($\tilde a(t) \neq a(t)$), then the second approach is needed to minimize the sample variance.  We have implemented both methods.\footnote{The code to generate the initial conditions for the separate universe cosmology given the fiducial cosmology, $\delta_{L0}$, and the box size choice is available upon request.}  
However, the matched-comoving-scale is appropriate for the application presented in this Letter (see \refsec{response}), and we therefore use separate universe simulations where the box size is given by $\tilde L = L \tilde h / h$. We choose $L= 500$ in units of $h^{-1}~{\rm Mpc}$, which is large enough to ensure that the growth of the largest mode in the box is very close to the linear prediction. This allows us to test our simulation set-up with perturbation theory predictions. 

We use Gadget-2 \citep{springel:2005} to carry out the N-body simulations. 
For the fiducial cosmology, we adopt a flat $\Lambda$CDM cosmology with $\Omega_m=0.27$ and $h=0.7$. Further parameters needed for the input power spectrum are the physical baryon fraction $\Omega_b h^2=0.023$, the spectral index $n_s=0.95$, and the amplitude of the primordial curvature power spectrum  $\mathcal A_s=2.2\times10^{-9}$.
In addition to the fiducial cosmology, we simulate  22 separate universe cosmologies sampling linear overdensities $\delta_{L0}$ between $-1$ and 1: 
$\pm( 0.01$, $0.02$, $ 0.05$, $ 0.07$, $ 0.1$, $ 0.2$, $ 0.3$, $ 0.4$, $ 0.5$, $ 0.7$, $1)$.
In these modified cosmologies $\tilde h$ decreases from 0.883  to 0.447 with increasing $\delta_{L0}$ and the curvature fraction $\tilde\Omega_K$ ranges from 0.37 to $-2.45$. In contrast, the physical fractions $\tilde\Omega_m\tilde h^2=0.1323$, $\tilde\Omega_\Lambda\tilde h^2=0.3577$, and $\Omega_b h^2=0.023$  as well as $n_s$  and $\mathcal A_s$ remain unchanged.
For each cosmology, we run 16 realization of the Gaussian random field and $256^3$ particles each.  
For one set of simulations, we increase the mass resolution by a factor of 8, i.e., $512^3$ particles, for use as a convergence test.  
All simulations are started at $z_i=49$.  

\vspace{-0.4cm}
\section{Power spectrum response}
\label{sec:response}

In this section we use the separate universe simulations to measure the growth-only response functions of the matter power spectrum to the linear overdensity $\delta_L(t)=\delta_{L0}D(t)/D(t_0)$. 
Let us denote the power spectrum of the separate universe cosmology by
$\tilde P(\tilde k,\delta_{L})$, where the power spectrum is measured as
a function of the comoving wavenumber and with respect to the mean density of the respective modified cosmology. For the fiducial cosmology, we have $\tilde P(\tilde k,\delta_{L}=0)=P(k)$. 
The growth-only response, which does not include effects due to the dilation of the physical scale nor to the change in the reference density, is defined by the derivatives of the power spectrum with respect to the mean overdensity at the same \emph{comoving} scale:
\be
G_n(k)=\frac{1}{ P(k)}\frac{d^n \tilde P(\tilde k, \delta_{L})}{d\delta_{L}^n}\bigg|_{\tilde k} \,,
\label{eq:Gndef}
\ee
evaluated at $\delta_L=0$.
As described in \cite{response}, the \emph{full} $n$-th order response of the power spectrum describes a certain
angle-averaged squeezed limit of the fully non-linear connected $n+2$-point 
function due to non-linear gravitational evolution, i.e., of the bispectrum
for $n=1$ and the trispectrum for $n=2$.  
Moreover, the response can be
measured in surveys via the position-dependent power spectrum technique
of \cite{posdeppk}.  

In this Letter, we restrict the analysis to the growth-only response, since this is the only contribution to the full response for which one actually needs to run separate universe simulations.  The growth-only response functions and the non-linear matter power spectrum of the fiducial cosmology are sufficient to compute the full response functions at all orders \citep{response}.

\begin{figure}
\includegraphics[clip=true,trim= 0.3cm 0.cm 0.7cm 0cm, angle=-90,width=80mm]{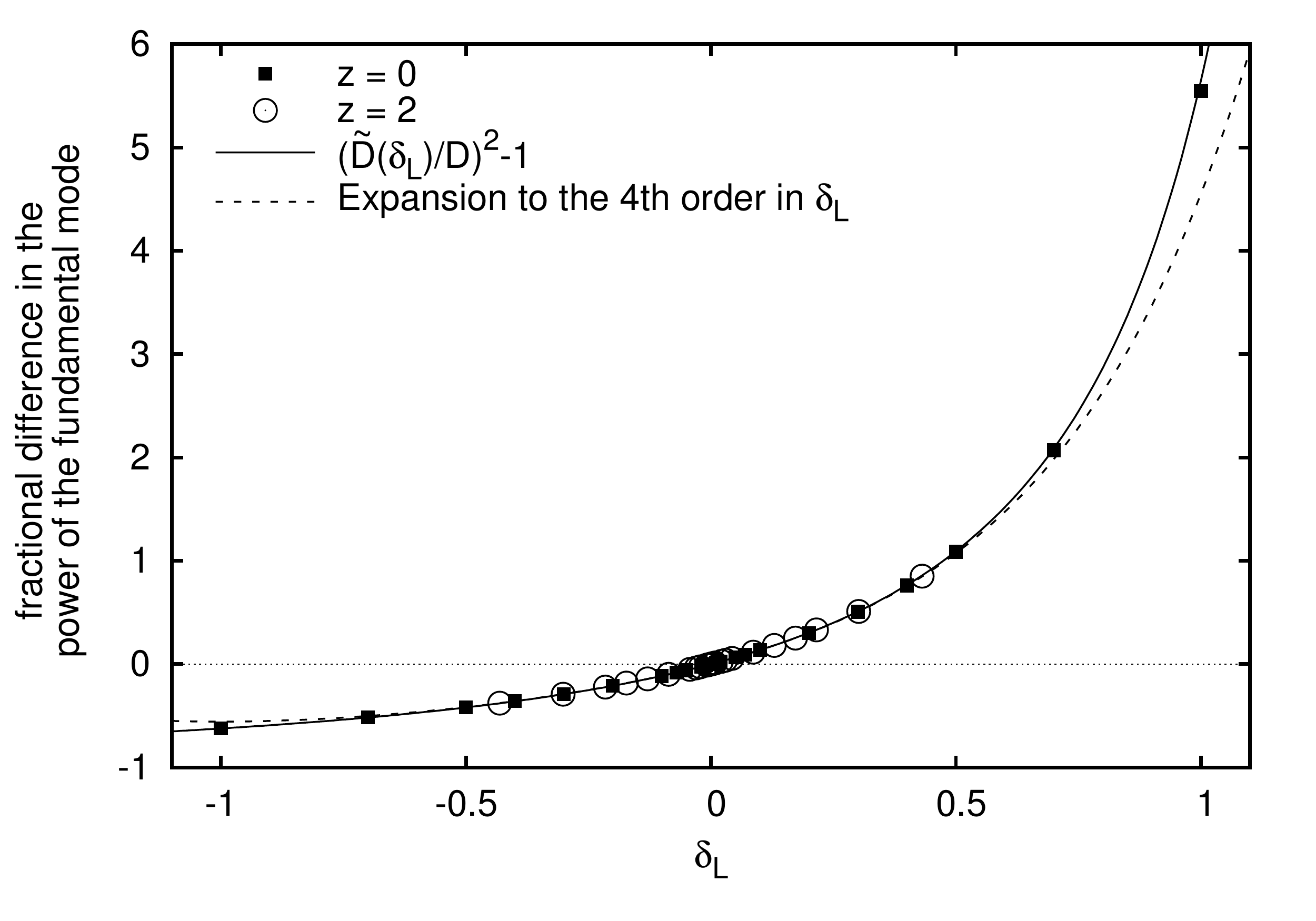}
\caption{The fractional difference in the power of the largest scale mode ($k_f\approx 0.01\ h\,{\rm Mpc}^{-1}$) measured from the separate universe simulations. The solid line corresponds to linear theory. The dashed line shows the predictions using the expansion given in \refeq{expansion}.}
\label{fig:delta}
\end{figure}

\begin{figure*}
\includegraphics[clip=true,trim= 1.2cm 1.cm 3.3cm 0.9cm, angle=-90,width=57mm]{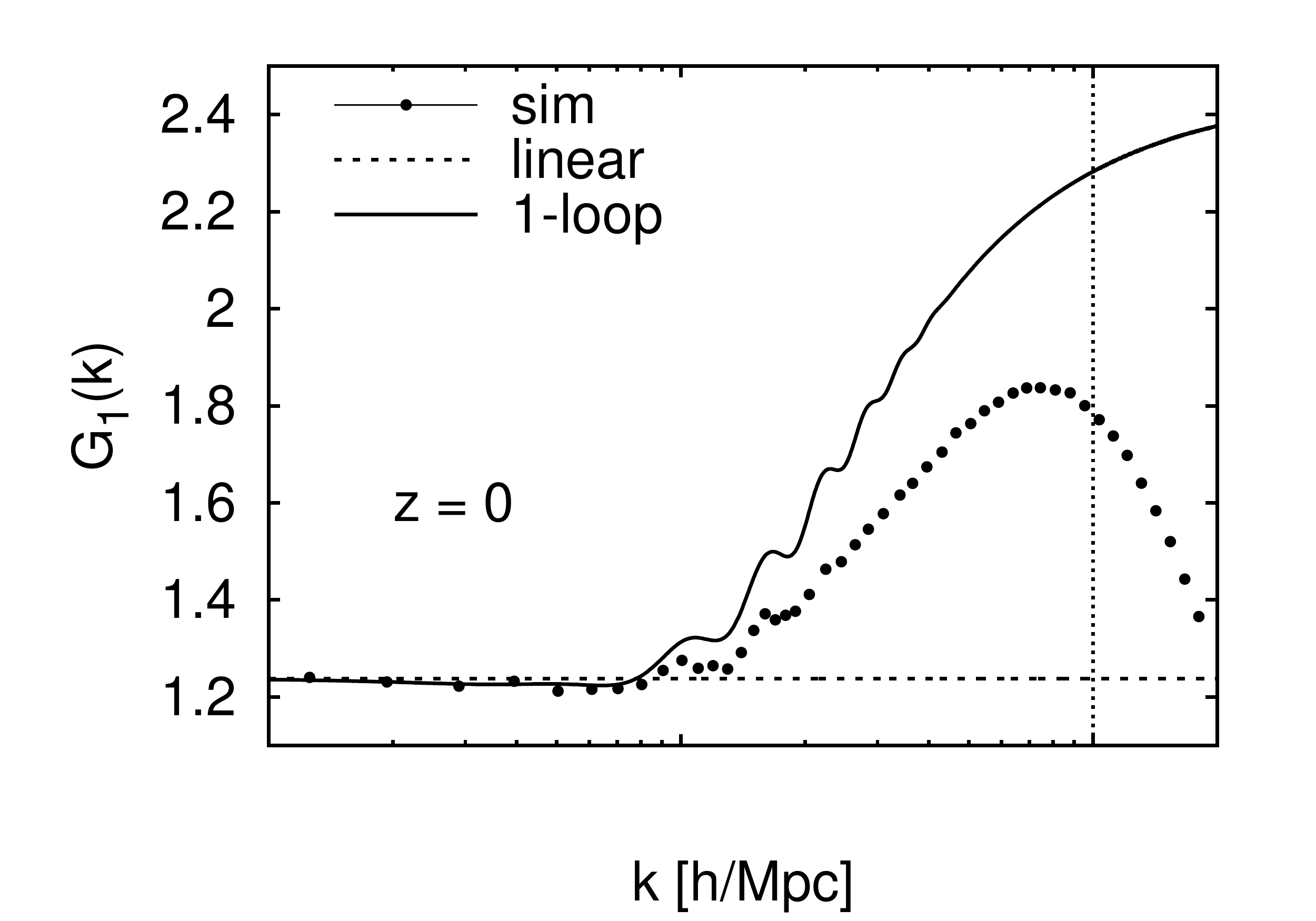}
\includegraphics[clip=true,trim= 1.2cm 1.cm 3.3cm 0.9cm,angle=-90,width=57mm]{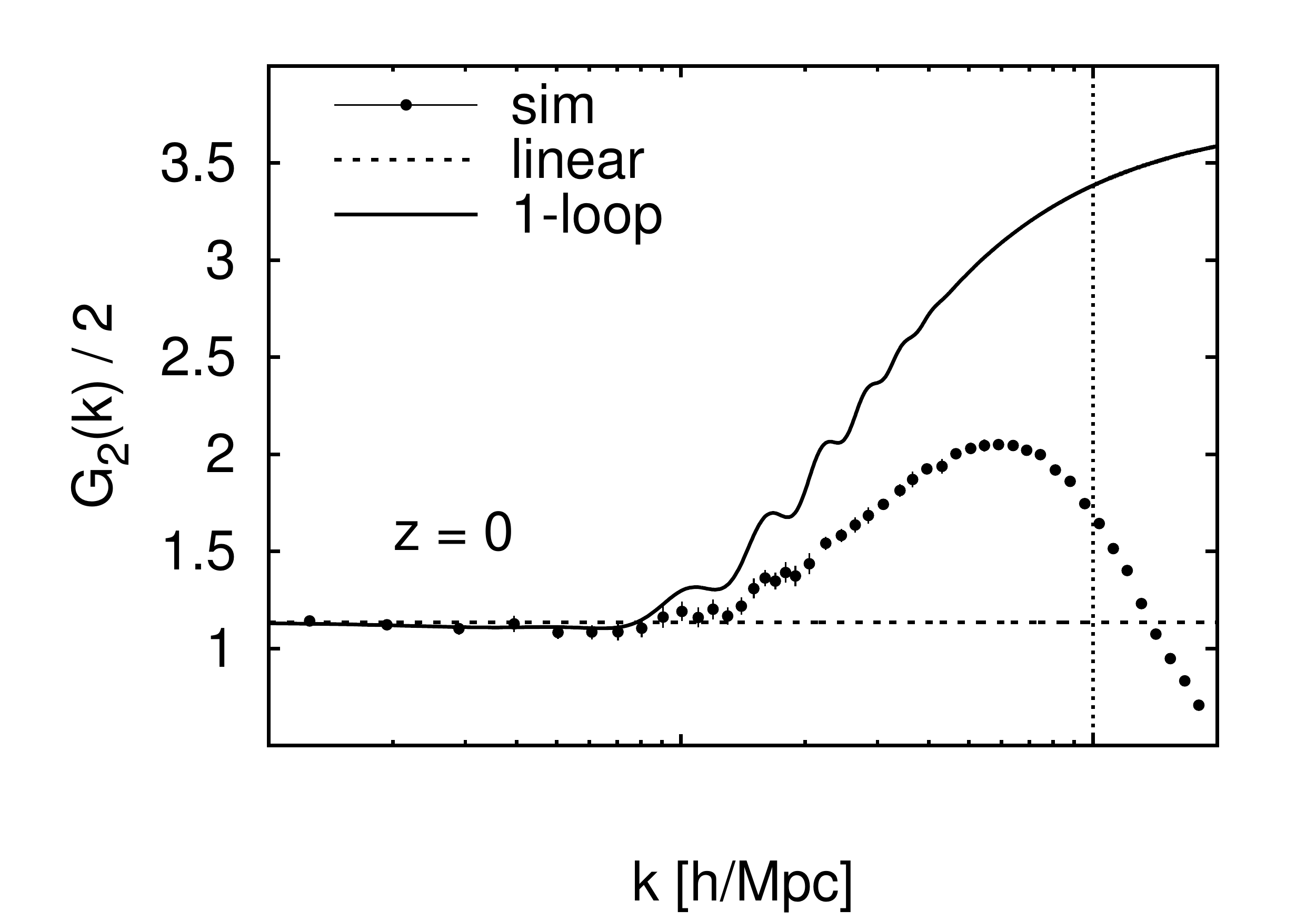}
\includegraphics[clip=true,trim= 1.2cm 1.cm 3.3cm 0.9cm,angle=-90,width=57mm]{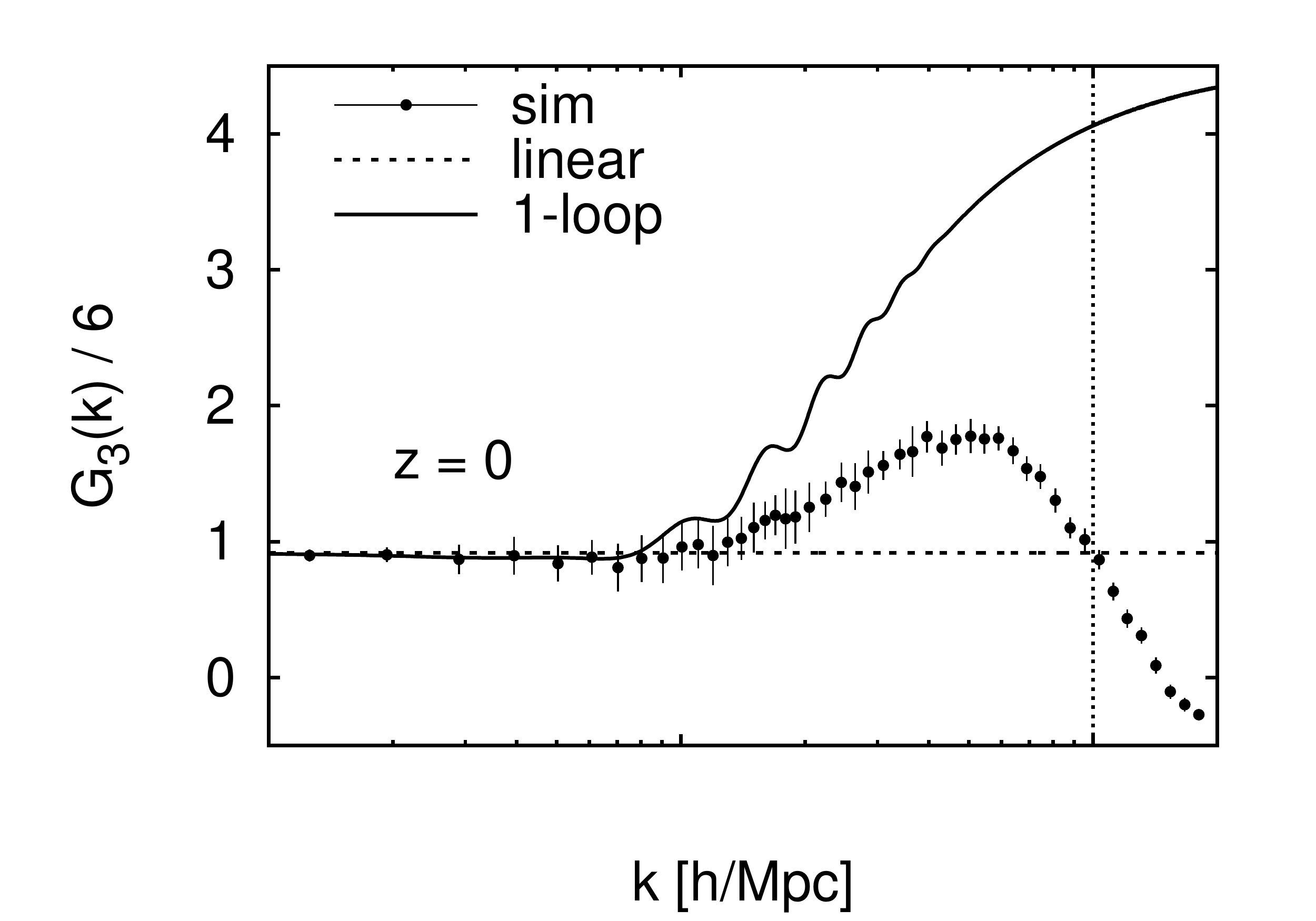}
\includegraphics[clip=true,trim= 0.7cm 1.cm 0.2cm 0.9cm, angle=-90,width=57mm]{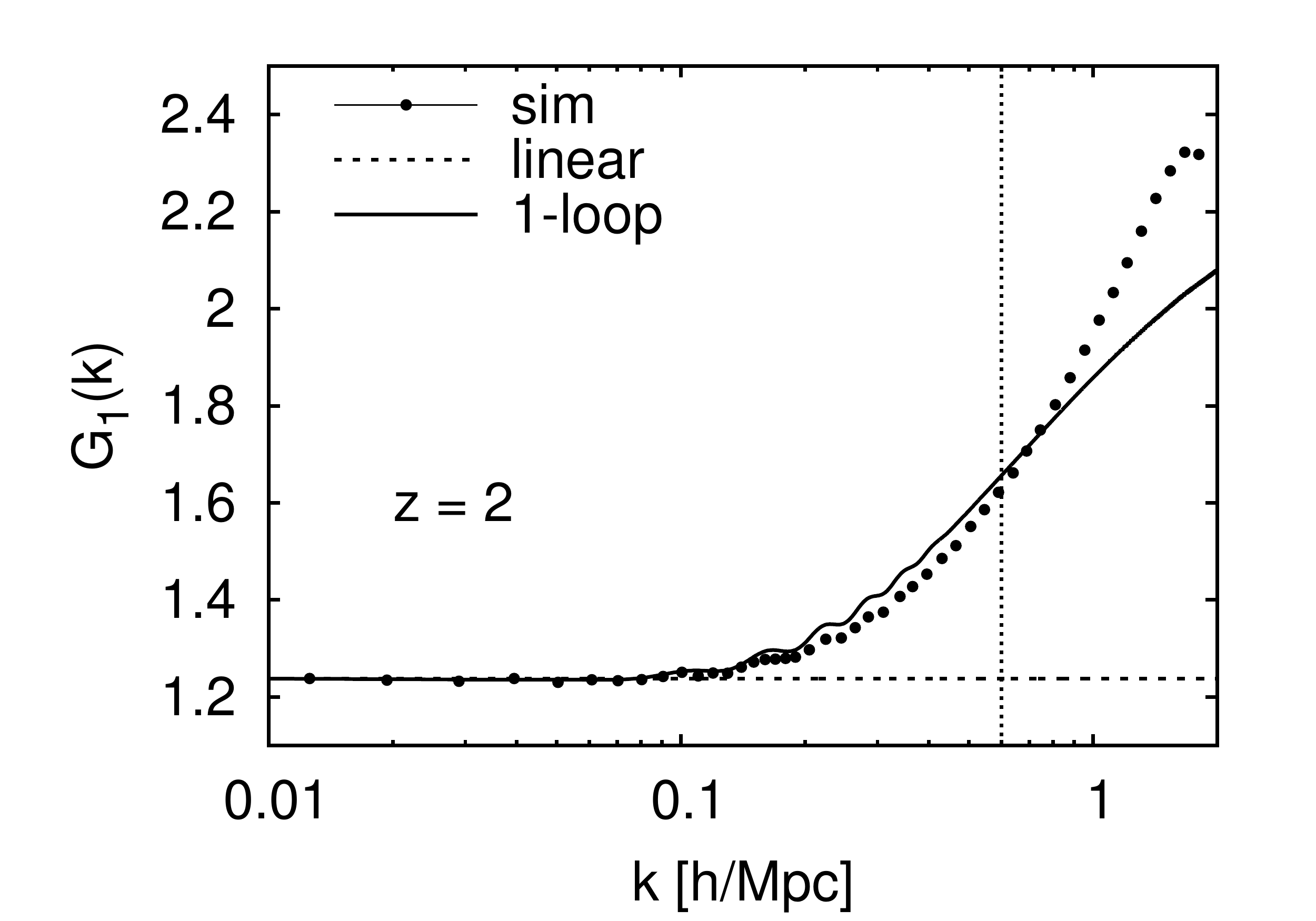}
\includegraphics[clip=true,trim= 0.7cm 1.cm 0.2cm 0.9cm,angle=-90,width=57mm]{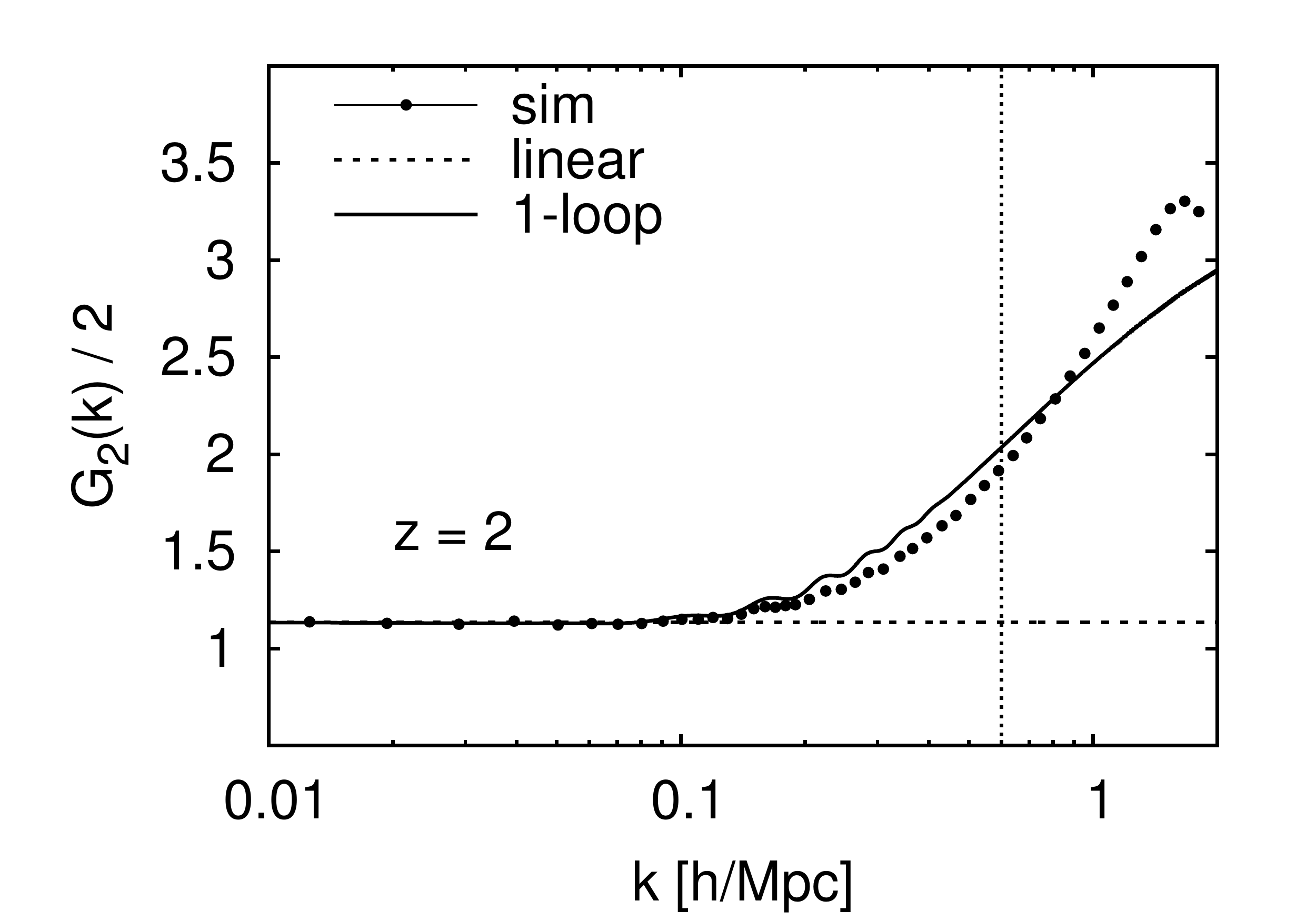}
\includegraphics[clip=true,trim= 0.7cm 1.cm 0.2cm 0.9cm,angle=-90,width=57mm]{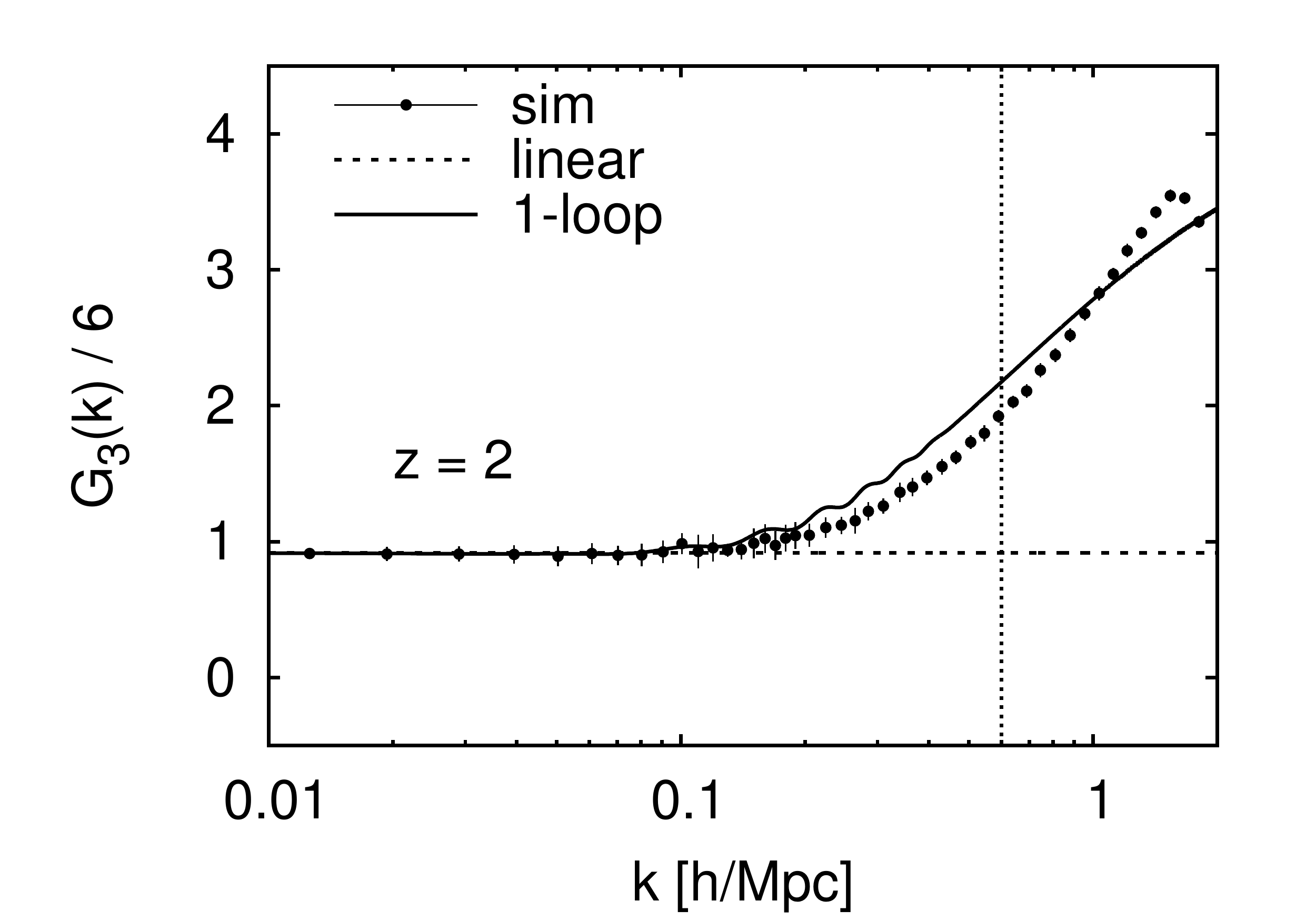}
 \caption{The first three growth-only response functions of the power spectrum [\refeq{Gndef}] measured from the separate universe simulations at $z=0$ (top) and $z=2$ (bottom). 
 The error bars show the statistical error derived by Jackknife
 sampling. For data points without error bars, the statistical error is
 smaller than the size of a dot.
 The vertical dotted line marks the maximum $k$ value for which the mass resolution of the simulations is sufficient to ensure that the results are converged to the 1-percent level. The dashed and solid lines show the analytic predictions using \refeq{expansion} in combination with the linear and 1-loop power spectrum, respectively.}
  \label{fig:response}
\end{figure*}

\refFig{delta} shows the fractional difference in the power of the fundamental mode of the box ($k_f\approx 0.01\ h\,{\rm Mpc}^{-1}$) with respect to the fiducial cosmology at $z=0$, as a function of the linear overdensity $\delta_{L}$, i.e.,~we show $\tilde P(\tilde k_f,\delta_{L})/P(k_f)-1$.\footnote{Note that the wavenumber and the power spectrum of the fiducial and modified cosmology are measured using the same units, namely the ones of the fiducial cosmology: $h~{\rm Mpc}^{-1}$ and $(h^{-1}~{\rm Mpc})^3$, respectively.}
The fundamental mode is sufficiently linear that the results should be
described well by the fractional difference in the linear growth:
$\tilde D^2(t)/D^2(t)-1$. The solid line shows the results of computing
numerically the linear growth for the separate universe cosmologies. As
expected the fractional growth measured on the largest scale of the
separate universe simulations is modelled very well by linear theory.
The dashed line shows the analytical prediction using the modification of the linear growth derived in an Einstein-de Sitter cosmology up to fourth order in $\delta_L$ \citep{response}: 
\ba
\tilde D(t)= D(t)\Big[& 1+\frac{13}{21}\delta_{L}(t)+\frac{71}{189}\delta_L^2(t)+\frac{29609}{130977}\delta_L^3(t)\vs 
&+\frac{691858}{5108103}\delta_L^4(t)+\mathcal{O}(\delta_L^5(t))\Big]\,.
\label{eq:expansion}
\ea
This formula agrees with the numerical result up to $\delta_L\approx 0.7$.

Next we measure the growth-only response functions as a function of $k$ by fitting a polynomial in $\delta_{L}$ to $\tilde P(\tilde k,\delta_{L})/P(k)-1$ measured from the simulations. 
The best-fit coefficients corresponding to the linear, 2nd-order and 3rd-order response functions are shown as data points in \reffig{response}. 
For the fit, we only include cosmologies with $|\delta_L| \leq 0.5$ and use a polynomial with degree 6 to be unbiased from higher-order response functions.
The error bars are derived by Jackknife sampling of the 16 realizations. The statistical error for $G_1$, $G_2$, and $G_3$ at $z=0$ is at the sub-percent, few-percent, and $\sim15\%$ level, respectively, and a factor of $\sim 2$ lower at $z=2$.  This impressively low level of noise achieved from a fairly small set of
simulations illustrates the power of the ``separate universe'' method.  
A simple analytic model can be derived by substituting the derivatives with respect to the mean overdensity $\frac{d}{d\delta_L}$ by derivatives with respect to the growth function $\frac{dD}{d\delta_L}\frac{d}{dD}$ and using the expansion given in \refeq{expansion}.
The dashed and solid lines show the predictions of this model using the
linear and 1-loop power spectrum, respectively. 
As expected, the range of validity of the 1-loop prediction significantly exceeds that of linear theory, and also increases with redshift.

\vspace{-0.4cm}
\section{Conclusions}
\label{sec:conclusion}

In the separate universe approach, the effect of a
long-wavelength overdensity $\delta_{L0}$ in the fiducial cosmology is
fully absorbed in a redefinition of the cosmology. This alternative
physical picture allows for a direct and self-consistent simulation of
this situation using standard cosmological N-body codes. Previously,
this approach was worked out to the lowest order in $\delta_{L0}$ \citep{mcdonald:2003,sirko:2005,li/hu/takada:2014} and used to compute the linear response function of the power spectrum \citep{li/hu/takada:2014}. 
Here, we extended the separate universe approach to all orders in the overdensity $\delta_{L0}$ 
and also pointed out the validity of the approach for certain
DE models beyond $\Lambda$CDM.    
We explicitly derived the mapping from the cosmological parameters of the fiducial cosmology to the separate universe cosmology as a function of $\delta_{L0}$.  
This enables one to run standard N-body codes also for larger
$\delta_{L0}$ and thereby study effects which are non-linear in
$\delta_{L0}$, while keeping all the other quantities fixed, especially the realization of the initial density field. 

We envision that this technique will be useful for a wide range of cosmological questions. In this Letter, in order to demonstrate and validate the approach we applied it---as a simple example---to measure the growth-only response of the matter power spectrum.  On scales for which the linear and 1-loop predictions are expected to be good approximations, we found excellent agreement between the simulation results and the theory.  A detailed modelling of these results including the non-linear regime and the physical interpretation will be presented together with the \emph{full} response of the matter power spectrum in \citet{response}.  Our results, obtained from a relatively small set of simulations, show exceedingly small error bars, at the sub-percent level in case of the linear response.  
This fact illustrates the power of this method, which can be applied in a 
broad range of contexts including more computationally intensive simulations
(e.g., including gas physics) in the case of which simulating large volumes 
is not feasible.  Furthermore, the isolation of the effect of a uniform 
density perturbation, excluding tidal effects as well as higher derivative
contributions, makes this approach ideally suited for robust measurements of
the local bias parameters of any given simulated tracer.  We plan to apply
this to dark matter halos in the future. 

\label{lastpage}

\vspace{-0.4cm}
\bibliography{references}

\end{document}